\documentclass[reprint,amsmath,amssymb,aps]{revtex4-2}

\usepackage{graphicx}
\usepackage{dcolumn}
\usepackage{bm}
\usepackage{hyperref}
\usepackage[mathlines]{lineno}
\nolinenumbers\relax 
\usepackage{natbib}
\usepackage{textcomp}
\usepackage{color}

\begin{document}
\bibliographystyle{apsrev4-2}
\preprint{APS/123-QED}

\title{Discovering dependencies in complex physical systems using Neural Networks}

\author{Sachin Kasture}
\email{sachinkasture84@gmail.com}
 \affiliation{OptoAI, 2805DL, Gouda, The Netherlands}
\date{\today}

\begin{abstract}
In todays age of data, discovering relationships between different variables is an interesting and a challenging problem. This problem becomes even more critical with regards to complex dynamical systems like weather forecasting and econometric models, which can show highly non-linear behaviour. A method based on mutual information and deep neural networks is proposed as a versatile framework for discovering non-linear relationships ranging from functional dependencies to causality. We demonstrate the application of this method to actual multivariable non-linear dynamical systems. We also show that this method can find relationships even for datasets with small number of datapoints, as is often the case with empirical data. 
\end{abstract}

\maketitle


\section{Introduction}
Finding relationships between different variables in large datasets \cite{reshef2011,marbach2010,brunton2016} is an important problem that has ramifications in fields ranging from environmental science to economics and genetic networks. Understanding what variables affect a certain quantity becomes increasingly challenging when these relationships are highly non-linear, like those occurring in dynamical systems with several variables. Quite often in a large dataset with several variables, only a few variables maybe significantly affecting the target variable and identifying these variables is first vital step in exploring these dependencies in more detail. 

Several methods exist which can help find dependencies and correlations between variables. However most of these methods are good at detecting a certain class of functions while they fail for others. There are some methods which are quite good at detecting functional dependencies between 2 variables \cite{reshef2011,dembo2001}, they have however not been demonstrated in a multi-variable scenario where a target variable depends on several input variables. Finding functional dependencies has been a topic explored extensively in context of relational databases\cite{liu2012,huhtala1999}. However these methods rely on finding exact functional relationships by finding all attributes which have a one to one or one to many relationship with a certain column Y. But this approach does not work well for small databases which are just a sample of the true distribution as in these cases one to one relations are more likely to occur. Also in such cases, it is difficult to reliably find the smallest subset of variables which are sufficient to describe Y. These methods do not offer any control over what kind of functional relationships maybe considered intuitively as good or interesting candidates. Also, these methods do not provide any kind of score to evaluate functional dependencies.

In this paper, we use Neural networks as devices to model nonlinear behavior and find complex non-linear relationships. Especially deep neural networks (DNN) which consist of more than 1 hidden layer are excellent candidates for efficiently modelling multi-variable non-linear polynomial functions with small number of neurons \cite{lin2017,rolnick2018}. Additionally a regularization mechanism allows us to control the complexity of the model we wish to consider \cite{tibshirani1996}. Neural networks have been used recently to discover physical concepts, identify phase transitions and design quantum experiments\cite{iten2020,rem2019,melnikov2018}. To help find dependencies, we use an DNN based autoencoder architecture which consists of an encoder-decoder pair. The encoder maps the input space to a latent space, while the decoder maps the latent space to the output space. This architecture has been used, amongst other applications, for non-linear Principle Component analysis (PCA) where the goal is to find a compressed representation of data \cite{hinton2006}. As such the input and the output of the autoencoder is conventionally the same. 
In our method the input will be $X$, which is the set of input features and $Y$ is the target feature or the set of features. We then use compression of mutual information in the latent space to derive a loss function which can be minimized to find the smallest set of features in $X$ which can be used to reliably reconstruct $Y$. The loss function can be used to assign a score to compare the functional dependencies on different set of input parameters.We then demonstrate this method to find dependencies in chaotic dynamical systems. Also we show that this method can be used to find non-linear causal connections in the Grangier sense for chaotic systems \cite{detto2012,runge2012,ma2015}, even for a small dataset of 100 samples.
\begin{figure*}[t]\centering
\includegraphics[scale=0.2,keepaspectratio=true]{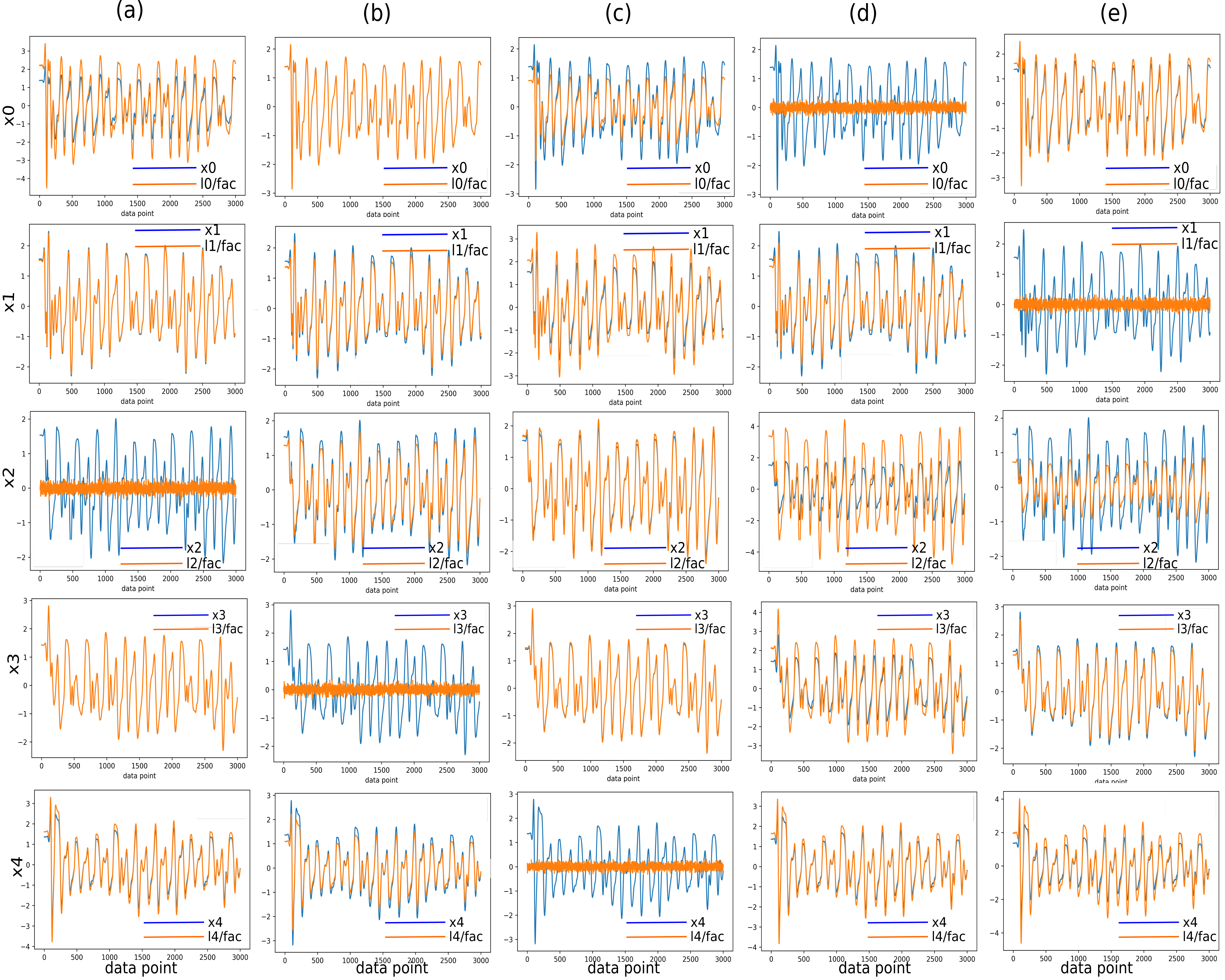}
\caption{\label{fig1} Plot shows comparison between $x_i$ and the corresponding scaled version of $l_i$ for (a)-(d) different values of $y_i=dx_i/dt$ for equation 17. In the plots where $l_i$ is essentially noise, information from the corresponding $x_i$ is not used to reconstruct $y_i$ using the decoder. $fac$ is a scaling factor chosen so that $x_i$ and $l_i/fac$ are comparable}
\end{figure*}
\section{Theory}
We now derive a loss function using the information bottleneck method \cite{tishby2000} based on the fact that the latent intermediate layer can be used to extract only relevant information from $X$ and used to reconstruct $Y$. We represent this latent representation by $L$. We also now assume a Markov chain $Y\rightarrow X\rightarrow\ L$. This means $P(Y|X,L)=P(Y|X)$. This is because $X,Y$ correspond to observed ground truth data.We now use the fact that we want to extract only relevant information from $X$ which can reconstruct $Y$. We use Shannon mutual information to quantify this information \cite{tishby2000,giannella2004}. Therefore want to maximize the quantity $I(L,Y)-\lambda_{enc} I(L,X)$. The first term and the second term describe the capacity of the encoder and the decoder respectively with $\lambda_{enc}$ determining the relative weight between the two terms. We can write $I(L,Y)$ as:
\begin{equation}
\begin{split}
    &I(L,Y) = \int dydl p(y,l)log\frac{p(y|l)}{p(y)}\\
    & = \int dl p(l)\int dy p(y|l)log(p(y|l) + H(Y)\\
\end{split}
\end{equation}
where $H(Y)$ is the Shannon entropy. We neglect $H(Y)$ since it is fixed by the data.
Since it is very difficult to calculate $p(y|l)$, we can approximate it by another analytic function $\phi(y|l)$. Using the fact that the KL divergence which measures the `distance' between 2 probability distributions is always non-negative:
\begin{equation}
\begin{split}
    &KL(p(y|l),\phi(y|l)) \ge 0 \\
&\implies \int dy p(y|l) log p(y|l) \ge 
\int dy p(y|l) log \phi(y|l)\\
\end{split}
\end{equation}
we can write
\begin{equation}
    I(L,Y) \ge \int dy dl p(y,l)log\phi(y|l)
\end{equation}
We can now choose an appropriate function for $\phi(y|l)$ which allows us to derive a suitable loss function as well as allows us to tune the complexity of the decoder. The output of the decoder is given by  $\theta_{dec}(l)$ which describes the composite function of the decoder neural network which acts on the latent variable $l$. To also include an additional L1 \cite{tibshirani1996}regulation parameter which helps restrict the magnitude of the weights in the decoder neural network, we use the following function for $\phi(y|l)$
\begin{equation}
    \phi(y|l) = e^{-(\theta_{dec}(l)-y)^2/\sigma_{dec}^2-\lambda_{dec}(|\theta_{d1}|+|\theta_{d2}|+..)}
\end{equation}
where $\theta_{d1},\theta_{d2}..$ etc. are weights of different neurons in the decoder network. Therefore we can write
\begin{equation}
\begin{split}
    I(L,Y)&\ge-\int dydlp(y,l)[\frac{(\theta_{dec}(l)-y)^2}{\sigma_{dec}^2}\\
    &+\lambda_{dec}(|\theta_{d1}|+|\theta_{d2}|+..)]\\
\end{split}
\end{equation}
Now we use the fact that $p(y,l)=\int dx p(x,y,l)=\int dx p(l|x,y)p(x,y)$. Using the Markov chain condition, this can be written as $p(y,l)=\int dxp(l|x)p(x,y)$. Approximating $\int dxdy p(x,y)A(x,y) = (1/M)\sum_{k=1}^M A(x^k,y^k) $ where $M$ is the number of distinct data points, we can write
\begin{equation}
\begin{split}
    I(L,Y)&\ge-(1/M)\sum_{k=1}^M\int dlp(l|x)[\frac{(\theta_{dec}(l)-y^k)^2}{\sigma_{dec}^2}\\
    &+\lambda_{dec}(|\theta_{d1}|+|\theta_{d2}|+..)]\\
\end{split}
\end{equation}
Similarly we can define $I(L,X)$ as:
\begin{equation}
\begin{split}
I(L,X) &= \int dldx p(x,l)log\frac{p(l|x)}{p(l)}\\
&=\int dxdlp(x,l)logp(l|x) - \int dlp(l)logp(l)\\
\end{split}
\end{equation}
We now again use another analytical function $g(l)$ in place of $p(l)$ and use the result on positivity of KL divergence and get:
\begin{equation}
    \begin{split}
    I(L,X) &= \int dldx p(x,l)logp(l|x) - \int p(l)logp(l)\\
    &\le \int dxdl p(x,l)log\frac{p(l|x)}{g(l)}\\
    \end{split}
\end{equation}
For convenience we use a Gaussian function centred at 0. 
\begin{equation}
    g(l) = e^{-\sum_i l_i^2/\sigma_{enc}^2}
\end{equation}
where  $l=(l_1,l_2..)$ are different components of $l$ and $\sigma_{enc}$ is an adjustable parameter.
For $p(l|x)$ we can use:
\begin{equation}
    p(l|x) = \prod_i e^{-(l_i-W_ix_i)^2/\sigma_{enc}^2} 
\end{equation}
where $x=(x_1,x_2,..)$ This means we use a linear transformation from $X$ and add a independent Gaussian noise with variance $\sigma_{enc}^2$ and mean 0 to each component. We now plug in definitions 9,10 into equation 8 and obtain:
\begin{equation}
    I(L,X)\le \int dxdlp(x,l)loge^{-\sum_i W_ix_i(W_ix_i-2l_i)/\sigma_{enc}^2}
\end{equation}
Writing $p(x,l)=p(x)p(l|x)$ we can write the above equation as
\begin{equation}
\begin{split}
    I(L,X)&\le -\int dxdlp(x)\prod_i e^{-(l_i-W_ix_i)^2/\sigma_{enc}^2} \\
    &[\frac{\sum_i W_ix_i(W_ix_i-2l_i)}{\sigma_{enc}^2}]\\
\end{split}
\end{equation}
Using the approximation $\int dxp(x)A(x) = (1/M)\sum_{k=1}^MA(x^k)$, we can write
\begin{equation}
\begin{split}
    I(L,X)&\le-(1/M)\sum_{k=1}^M\int dl \prod_i e^{-(l_i-W_ix_i^k)^2/\sigma_{enc}^2} \\
    &[\frac{\sum_iW_ix_i^k(W_ix_i^k-2l_i)}{\sigma_{enc}^2}]\\
\end{split}
\end{equation}
Similarly substituting equation 10 into equation 6 and assuming $\sigma_{enc}^2$ to be small enough so that $e^{-(l_i-W_ix_i)^2/\sigma_{enc}^2}\approx\delta(l_i-W_ix_i)$we obtain:
\begin{equation}
\begin{split}
    I(L,Y)&-\lambda_{enc} I(L,X)\ge-(1/M)\sum_{k=1}^M[\frac{(\theta_{dec}(l)-y^k)^2}{\sigma_{dec}^2}+\\ 
    &\lambda_{dec}(|\theta_{d1}|+|\theta_{d2}|+..)+\lambda_{enc} \sum_i\frac{(W_ix_i^k)^2}{\sigma_{enc}^2}]\\
\end{split}
\end{equation}
\begin{figure}[t]\centering
\includegraphics[scale=0.3]{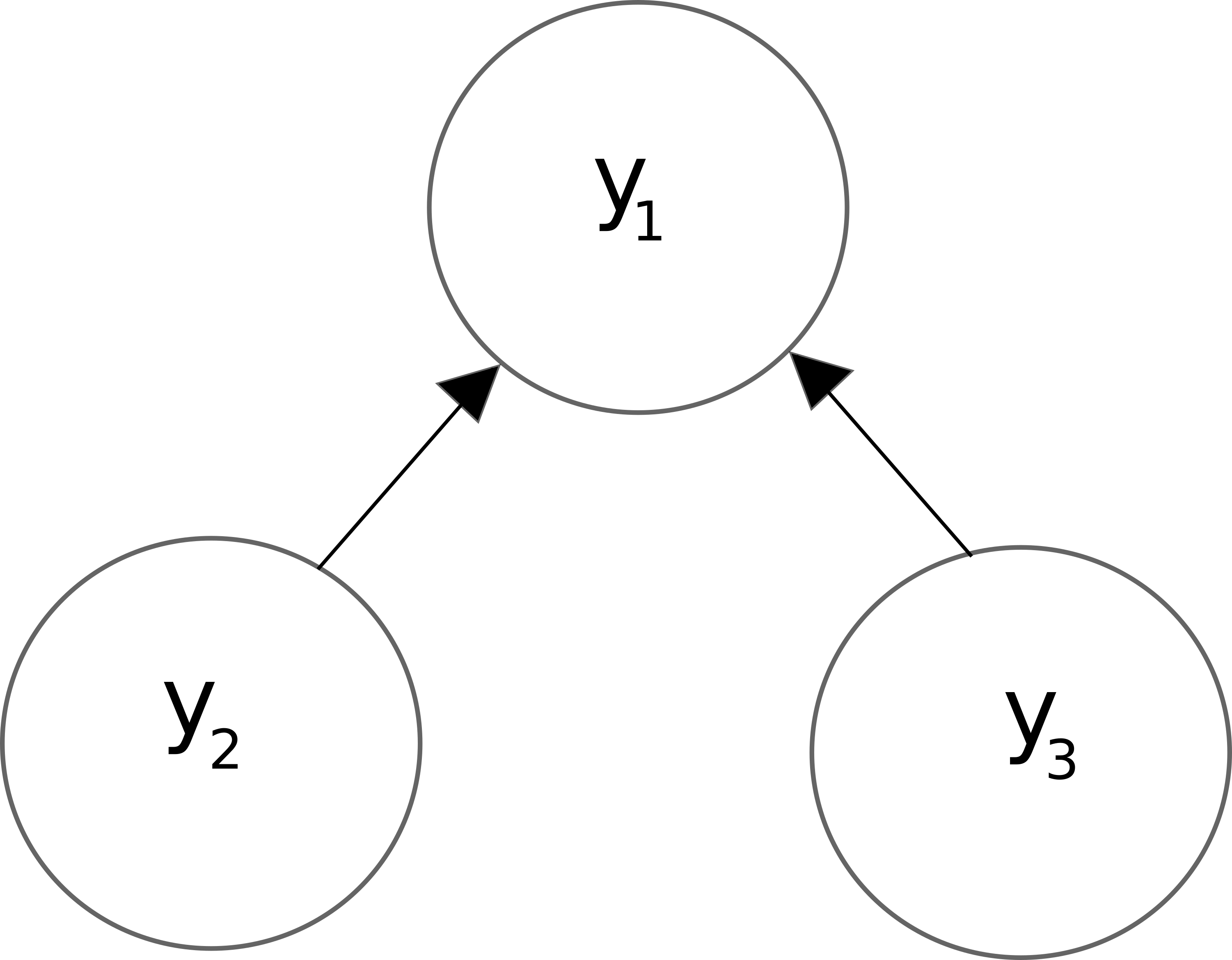}
\caption{\label{fig2} Plots shows the case of fan-in causality pattern for set of delay equations in equation 18 for set of $\xi_{ij}$ values used to obtain results in Figure 3}
\end{figure}
\begin{figure*}[t]\centering
\includegraphics[scale=0.3]{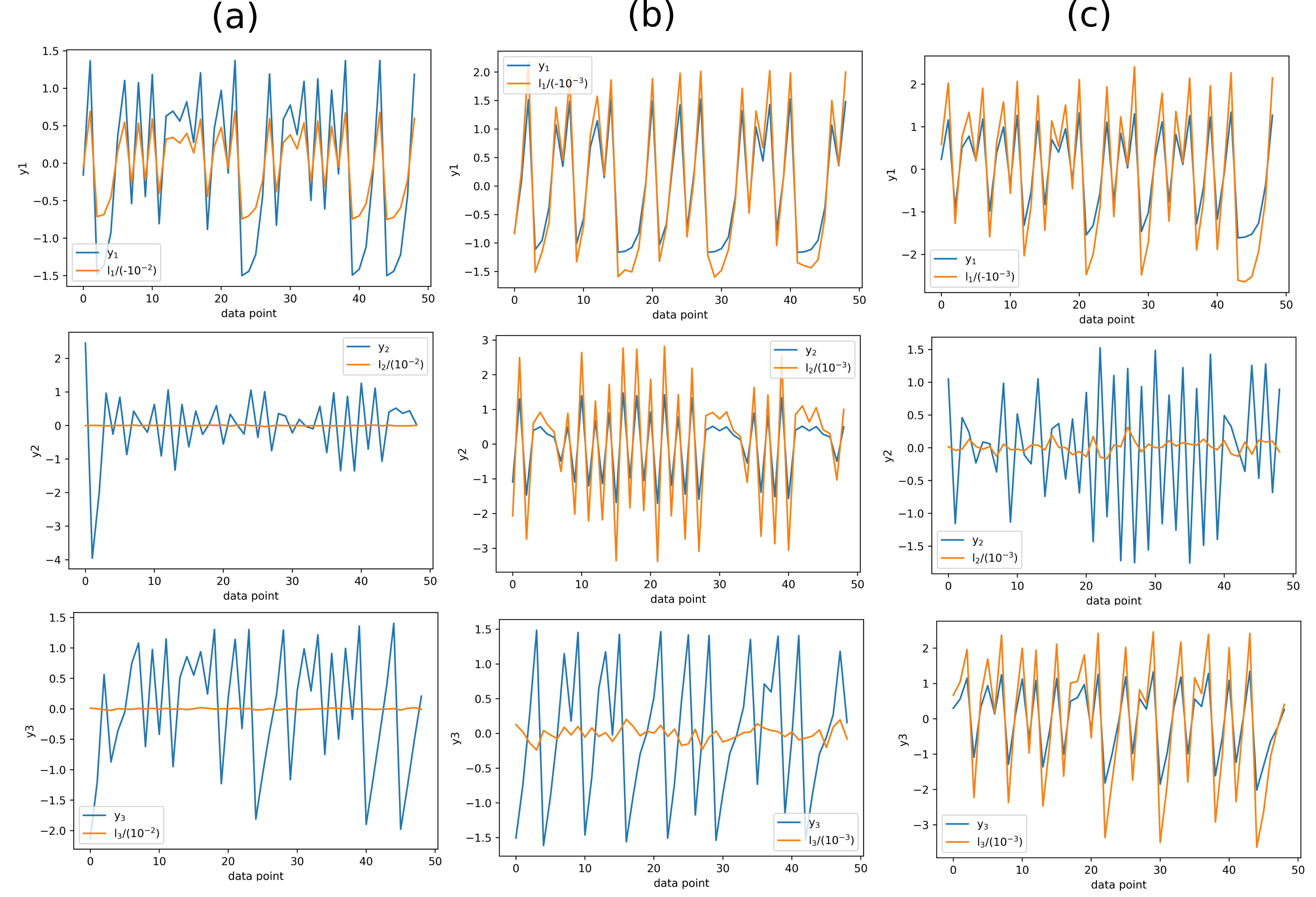}
\caption{\label{fig3} Plot shows comparison between $Y_i$ and the corresponding scaled version of $l_i$ for (a)-(c) different values of $y_i=Y_i$ for the set of delay equations 18. In the plots where $l_i$ is noise, information from the corresponding $x_i$ is not used to reconstruct $y_i$ using the decoder}
\end{figure*}
Therefore we can define a loss function to be minimized as 
\begin{equation}
\begin{split}
    \mathcal{L} &= (1/M)\sum_{k=1}^M[\frac{(\theta_{dec}(l)-y^k)^2}{\sigma_{dec}^2}+\\ 
    &\lambda_{dec}(|\theta_{d1}|+|\theta_{d2}|+..)+\lambda_{enc} \sum_i\frac{(W_ix_i^k)^2}{\sigma_{enc}^2}]\\
\end{split}
\end{equation}
We observe that the first term tries to minimize the least squares difference between $\theta_{dec}(l)$ and $y$ and the second term controls the size of the weights of the decoder which in turn controls the maximum degree polynomials the decoder NN can approximate. For the third term we see that as we increase the $\lambda_{inc}$, the NN will try to keep $(W_ix_i^k)^2$ small to keep the total loss function small. Assuming now that we standardize our data so that $x_i's$ on an average have similar magnitudes, we absorb it into $\lambda_{enc}$. The third term will now be smallest when only $W_i's$ corresponding to those $x_i's$ are non-zero, which are required to reproduce $Y$. Using this intution and the fact that term inside the summation over $i$ in equation 17 is always $\ge 0$, we can further simplify the loss function as
\begin{equation}
\begin{split}
    \mathcal{L} &= (1/M)\sum_{k=1}^M[\frac{(\theta_{dec}(l)-y^k)^2}{\sigma_{dec}^2}]+\\ 
    &\lambda_{dec}(|\theta_{d1}|+|\theta_{d2}|+..)+\lambda_{enc} \sum_i(|W_i|)\\
\end{split}
\end{equation}
where we have merged $\sigma_{enc}^2$ with $\lambda_{enc}$. 
This way we treat both the encoder and decoder weights on equal terms using L1 regularization. From a practical standpoint L1 is advantageous since it can shrink weights faster. 
\section{Application}
For further study we use a NN in which the encoder has 2 linear layers. This gives us a mapping $X\rightarrow L$. We then add Gaussian noise to the latent variables $l_i=l_i+N(0,\sigma^2_{enc})$. The latent code is then sent through a multilayer decoder network with non-linear activation functions to give the output $\theta_{dec}(l)$. We perform batch-normalization in between intermediate neural network layers \cite{ioffe2015}. This layers prevents change in data distributions between adjacent layers and allows neural network learning at a higher learning rate. We then minimize the loss function in equation 16 using Stochastic gradient descent with different batch sizes. We can tune the values of $\lambda_{enc},\lambda_{dec}$ (regularization parameters) to obtain as low values of loss function as possible. This choice of regularization parameters may also depend on our prior knowledge about the complexity of the system. The data is split into the training and validation set. The training data is used to build the model and validation set checks how well the model generalizes. The basic heuristic for tuning these parameters is as follows: after fixing the learning rate for the gradient descent, we first increase the value of $\lambda_{dec}$ which basically fixes the complexity of functions the decoder can simulate. We then increase the value of $\lambda_{enc}$ and look at the value of the mean square error and stop when the mean square error is as small as possible for both the training and the validation set. We now use this method to infer relationships in well known non-linear systems. We first consider a Lorenz96 non-linear system which is defined as:
\begin{equation}
    \frac{dx_i}{dt} = (x_{i+1} - x_{i-2}x_{i-1} - x_i + F)
\end{equation}
where $i$ goes from $1$ to $N$ where $N$ is the number of oscillators and $x_{N+1}=x_1$,$x_{-1}=x_{N-1}$, $x_0=x_N$. $F$ is the driving term and we choose $F=8$ where the system behaves in the chaotic regime. Figure 1 shows the results for N=5. We run N=5 times with each time $y=\frac{dx_i}{dt}$ for i from 1 to 5. We see that the latent representation $l_i$ is basically just the added Gaussian noise when the corresponding $y$ has no dependency on $l_i$. The number of data points was 3000 and learning rate was 0.0001 and values of $\lambda_{dec},\lambda_{enc}$ where 0 and 0.1 respectively. The training was run for 1000 epochs with a batch size of 300. 
\\ \indent
Next we apply NN to infer causal relationship in a set of non-linear delay equations. For this we look at the following set of equations:
\begin{equation}
    Y_{i}(t+1) = (\xi_{ii}-\sum_{j=1,2,3}(\xi_{jj}-\xi_{ij}Y_{j}(t)))Y_i(t)
\end{equation}
for i=1,2,3. We choose to choose parameters $\xi_{ij}$ which correspond to a fan-in pattern shown in Figure 2.  The values of $\xi$ are as follows $\xi_{11}=4,\xi_{22}=3,\xi_{33}=2,\xi_{31}=0.6,\xi_{32}=-0.6$. These parameters corresponds to a chaotic regime. In this case both $Y_2$ and $Y_3$ are causally driven by $Y_1$. A fan-in pattern is a good test because correlation based tests would falsely infer a causal relationship between $Y_2$ and $Y_3$ \cite{marbach2010}. To infer the causal relationships, we run the NN with $y=Y_i(t+1)$ and input $X=[Y_1(t),Y_2(t),Y_3(t)]$. From Figure 3 we can see that we are able to correctly infer the dependencies, even for a very small data-set of 50 points. The plots were obtained for a learning rate of 0.001 and $\lambda_{enc},\lambda_{dec}$ values of 0.1 and 0.005 respectively.The number of epochs was 1500 with a batch size of 32.
\begin{figure}[t]\centering
\includegraphics[scale=0.4]{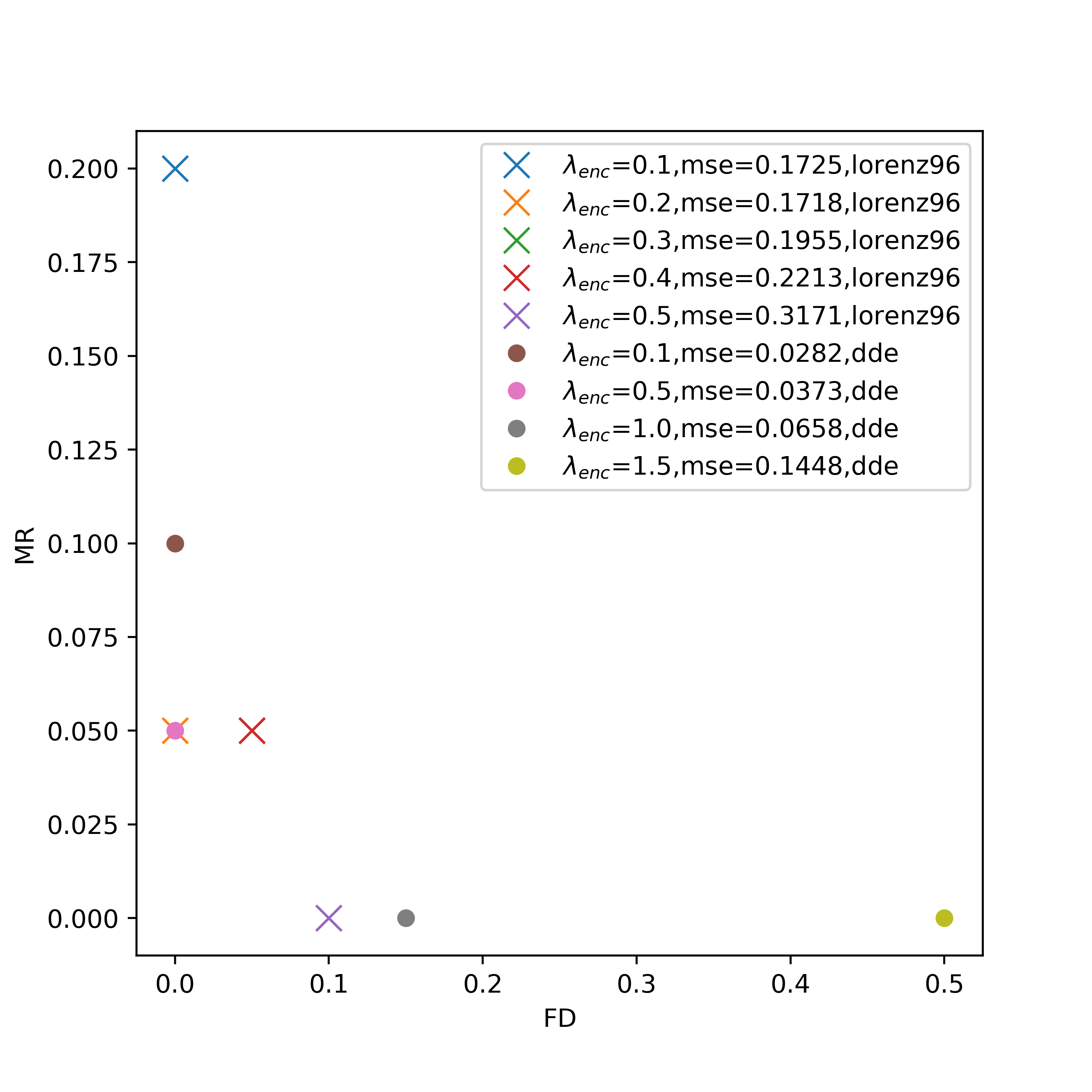}
\caption{\label{fig4} Plot shows the plot for FD vs MR for different values of $\lambda_{enc}$. The legend also mentions the non-linear system for the plotted data. `dde' stands for the delay difference equations in equation 18 }
\end{figure}
We also summarize the performance of this method using 2 metrics False discovery (FD) and Miss rate (MR) which are defined as:
\begin{equation}
\begin{split}
& FD = \frac{FP}{FP+TP}\\
& MR = \frac{FN}{FN+TP}\\
\end{split} 
\end{equation}
where FN, FP, TP are False negatives, false positives and true positives respectively. Here a positive means a certain variable has been discovered to be independent of the output. The negative means a variable has been discovered to be related to the output.This data is obtained by obtaining results over 20 independent runs of the model. For the Lorenz96 model, the best result is obtained with $\lambda_{enc}=0.2$ while for the set of equations 18, best results are obtained for $\lambda_{enc}=0.1$  
\section{Conclusion}
The proposed approach using NN is a versatile platform for inferring relationships, especially in complex non-linear systems. This is because NN are a powerful tool to model such non-linear functions. Even though it is difficult to infer the exact functional form using a NN, this method can help locate functional dependencies between variables in a multivariable system. These variables can then be probed more extensively to find the functional (or approximate functional) form of the relationships. Methods based on sparse regression have been used in the past to find functional relationships. However they rely on pre-knowledge of the set of basis functions to use for the regression. The proposed method has no such requirement and with a large enough NN, can simulate any complex non-linear function. Besides locating functional relationships, it can also help infer causal relationships in non-linear data as seen in the discussed example, where it correctly inferred causal relationship even for a small dataset of 50 samples.
\section{Acknowledgements}
The author would like to thank Akshatha Mohan for helpful comments and critical assessment of the manuscript.
\bibliography{NN_references}
\end{document}